\documentclass[letterpaper, 10 pt, journal, twoside]{ieeetran}

\IEEEoverridecommandlockouts                              

\usepackage[utf8]{inputenc}
\usepackage{amsmath,amssymb,amsfonts}
\usepackage{graphicx}
\usepackage{siunitx}
\usepackage{hyperref}
\hypersetup{hidelinks}
\usepackage{algorithm}
\usepackage{algorithmic}
\usepackage{multirow}
\usepackage{hyperref}
\usepackage{balance}
\usepackage{float}

\begin{document}

\title{\LARGE \bf
Distributionally Robust Acceleration Control Barrier Filter\\ for Efficient UAV Obstacle Avoidance
}

\author{Dnyandeep Mandaokar$^{1}$ and Bernhard Rinner$^{1}$
\thanks{Manuscript received: November, 29, 2025; Revised February, 11, 2026; Accepted March, 14, 2026.}
\thanks{This paper was recommended for publication by Editor Giuseppe Loianno upon evaluation of the Associate Editor and Reviewers' comments.} 
\thanks{$^{1}$Dnyandeep Mandaokar and Bernhard Rinner are with the Institute of Networked and Embedded Systems, University of Klagenfurt, 9020 Klagenfurt, Austria
        ({\tt\small dnyandeep.mandaokar@aau.at} and {\tt\small bernhard.rinner@aau.at}).}
\thanks{Digital Object Identifier (DOI): see top of this page.}}%

\markboth{IEEE Robotics and Automation Letters. Preprint Version. Accepted March, 2026}
{Mandaokar \MakeLowercase{\textit{et al.}}: Distributionally Robust Acceleration Control Barrier Filter for Efficient UAV Obstacle Avoidance}

\maketitle
\begin{abstract}

Dynamic obstacle avoidance (DOA) for unmanned aerial vehicles (UAVs) requires fast reaction under limited onboard resources. We introduce the distributionally robust acceleration control barrier function (DR-ACBF) as an efficient collision avoidance method maintaining safety regions. The method constructs a second-order control barrier function as linear half-space constraints on commanded acceleration. Latency, actuator limits, and obstacle accelerations are handled through an effective clearance that considers dynamics and delay. Uncertainty is mitigated using Cantelli tightening with per-obstacle risk. A DR-conditional value at risk (DR-CVaR) early trigger expands margins near violations to improve DOA. 
To meet real-time avoidance-control at 100\,Hz, we use fixed-time Gauss-Southwell projections instead of quadratic programs (QP). Simulation results show similar avoidance performance with 31\% lower computational load than QP and outperform the state-of-the-art baseline approaches. Experiments with Crazyflie UAVs demonstrate the feasibility of our approach.    

\end{abstract}


\section{Introduction}
\label{sec:INTRO}
\IEEEPARstart{U}{nmanned} aerial vehicles (UAVs) are increasingly deployed in day-to-day services that require rapid responses to dynamic obstacles. Applications include urban UAV logistics, disaster response, and inspection in cluttered environments where obstacles such as other UAVs, birds, or humans may appear or move unpredictably~\cite{Rinner_Computer2021}. To operate safely in these cases, UAVs must react fast to avoid collisions while maintaining stability. Dynamic obstacle avoidance (DOA) relies on high-rate perception and control and must handle perception uncertainty, latency, and limited computation under tight real-time constraints. Even a small perception-to-actuation latency can cause severe safety violations at high speeds~\cite{falanga2019ral}. In addition, vision and radio detection and ranging (RADAR) sensors produce high-rate measurements that further strain the capabilities of current DOA systems~\cite{mandaokar_widroit2025,falanga2019ral}.

Classical UAV sense-and-avoid approaches rely on trajectory optimization~\cite{chen2019robio,foam2021,rota_icra2024}, nonlinear model predictive control (NMPC)~\cite{lindqvist2020ral}, or reactive planners~\cite{lin2019fast3d,singletary2021iros}. While these methods can handle static or slowly moving obstacles, they are computationally demanding or rely on conservative safety margins, making them unsuitable for agile flights. Moreover, these optimization-based approaches often lack formal safety guarantees in the presence of fast and uncertain obstacle motion. Recently, control barrier functions (CBFs) emerged as a reliable framework to guarantee safety by ensuring forward invariance of a safe set~\cite{AdaptiveCBF2022,ames2017tac,ames2014acc} throughout the entire mission. Classical CBFs are often applied via quadratic programs (QPs) to enforce safety while minimally adjusting nominal control, enabling real-time collision avoidance for robots, manipulators, and UAVs. Similarly, the existence of higher-order CBF (HOCBF) extends CBFs to handle higher relative-degree systems~\cite{AdaptiveCBF2022}, which is crucial for UAVs with second-order dynamics. The VOCBF~\cite {VOCBF2025} encodes safety in velocity space. Moreover, the CBF is integrated with differentiable optimization-based methods to enable smoother control pipelines~\cite{dai2023ral}. Similarly, a probabilistic enumeration is introduced for stochastic risk allocation~\cite{ProbEnum2025}. Runtime adaptive safety filtering and robust formulations have further broadened the applicability of CBFs in robotics~\cite{SafetyFly2025,RobustCBF2025}.

Despite this progress, current CBF implementations face limitations for DOA in UAVs. Existing CBF formulations act at the position or velocity level~\cite{VOCBF2025,FixedWingCBF2025}, but do not provide closed-form acceleration-space constraints compatible with fast low-level thrust control. The HOCBFs do not directly address acceleration or thrust for UAV DOA~\cite{HOCBF1}. Although their constraints are solver-independent, online QP approaches can limit real-time use on embedded platforms~\cite{AdaptiveCBF2022}. The current CBF is insufficient to address latency and uncertainty in UAV DOA. The ISSf-CBF~\cite{ISSfCBF2023} and tube-based formulations~\cite{TubeCBF2023} consider robustness to input delay, but they are conservative and rely on iterative solvers. Probabilistic CBFs improve risk handling, but have not been adapted to acceleration-level constraints or integrated with hard real-time systems~\cite{ProbEnum2025,SafetyFly2025}. Similarly, CBF with chance-constraints emphasizes probabilistic risk margins for UAVs but remains velocity-centric~\cite{cccbf}. The QP-based CBF solvers and sampling-based planners are too slow for high-rate DOA control, particularly on resource-constrained UAV hardware~\cite{FixedWingCBF2025,RobustCBF2025,AdaptiveCBF2022}. Lightweight and computationally efficient MPC extensions, such as  TinyMPC, proving formal guarantees for fast reaction, are available for Crazyflies, but they do not quantify performance in high closing velocity DOA scenarios~\cite{tinyMPC}.

In this paper, we address the current CBF-QP limitation for DOA in UAVs with a \textit{distributionally robust acceleration control barrier function (DR-ACBF)}. DR-ACBF directly restricts UAV acceleration commands through safe, linear, half-space constraints. The half-space constraints specify the safe side of the projection plane for the UAV's acceleration, ensuring that the UAV remains within this region. 
We demonstrate that enforcing a linear acceleration-space constraint derived from the second-order CBF dynamics satisfies the original CBF inequality over a short horizon under bounded uncertainty and latency assumptions. Additionally, the perception-to-actuation latency and uncertainty are handled by incorporating clearance buffering~\cite{falanga2019ral} and probabilistic tightening based on Cantelli inequalities~\cite{Cantelli}, similar in spirit to stochastic CBFs~\cite{ProbEnum2025,SafetyFly2025} but specialized for acceleration space. Moreover, an early-warning trigger layer based on distributionally robust conditional value at risk (DR-CVaR)~\cite{cvar2023,drcvar2023,wasserstein} anticipates risk growth within the same maximum horizon $H_{max}$ and inflates safety margins before any violation becomes imminent. This trigger improves the robustness of ACBF by ensuring that subsequent ACBF enforcement remains feasible and effective under latency- and jerk-limited execution. We use fixed-time Gauss-Southwell projection~\cite{gauss,fastproj} instead of online QPs, enabling real-time control at a rate of 100\,Hz.

In contrast to existing CBF-driven collision avoidance frameworks, our DR-ACBF framework provides: (i) explicit acceleration-space tightening to stay safe, (ii) low computational cost via iterative projection, and (iii) probabilistic and latency-aware safety margins. We derive a closed-form line-of-sight (LoS) acceleration half-space with horizon-matched gains and a fixed-time projection enforcement with Gauss-Southwell. To the best of our knowledge, no prior works present an acceleration-space CBF filter with latency and risk tightening for UAV obstacle avoidance. The DR-ACBF achieves real-time, reliable UAV obstacle avoidance via acceleration-level CBFs, DR-CVaR trigger, and fixed-time projection, reducing command time by over 32\% while maintaining high safety and success rates as evaluated in a simulation study. Projection-based DR-ACBF reduces processing time by 31\% as compared to QP-based solvers. Experiments with Crazyflie UAVs confirm the feasibility of our approach.

\section{Problem Formulation}

\begin{figure}[thpb]
  \centering
  \includegraphics[width=\columnwidth]{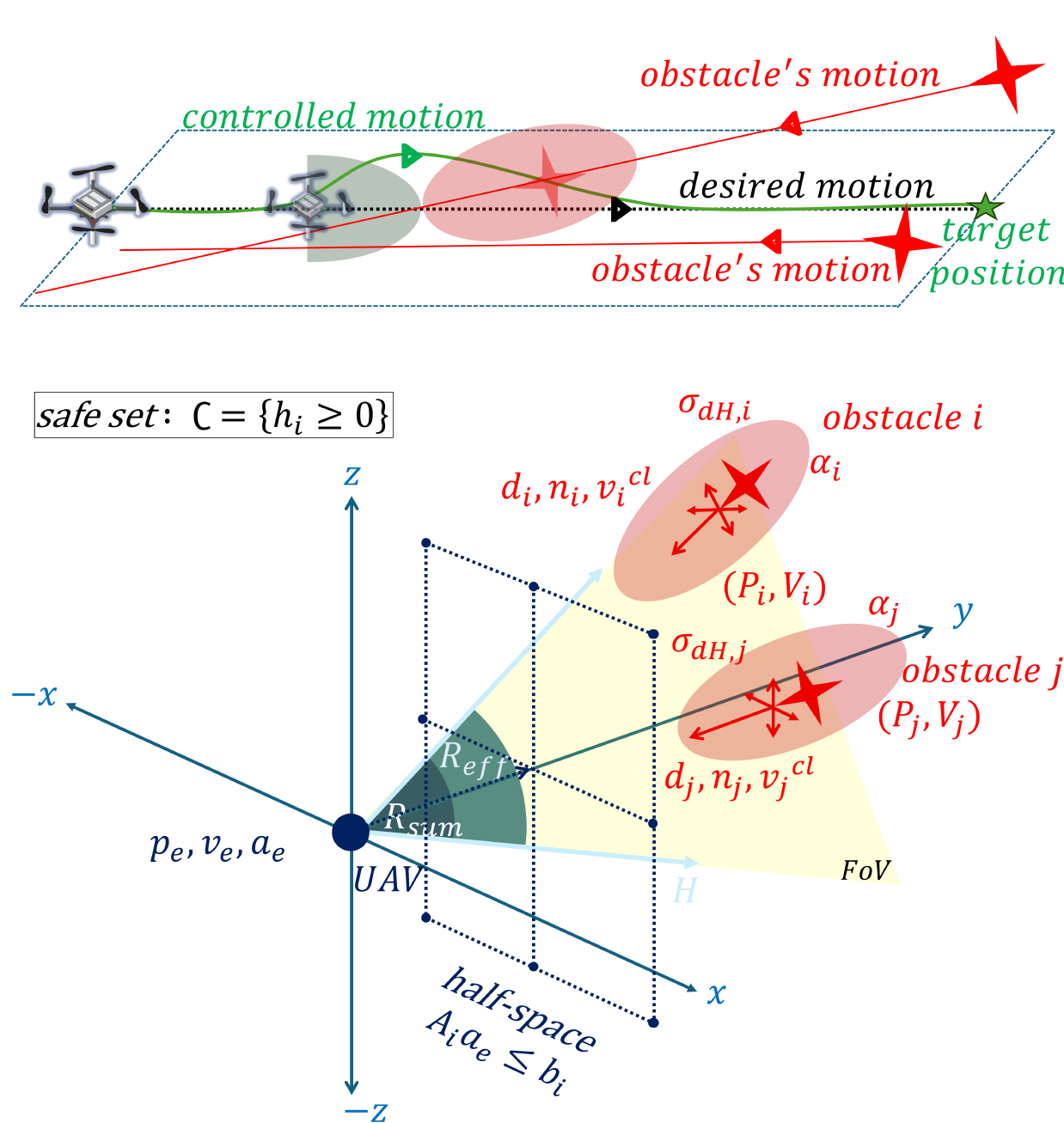}
  \caption{Illustration of the proposed DR-ACBF framework. 
The UAV avoids dynamic obstacles (top) by constraining its acceleration within distributionally robust half-spaces marked in grey. The UAV view (bottom) depicts the projected half-space plane towards the obstacles in FoV. The effective clearance $R_{\mathrm{eff}}$ accounts for latency and actuation limits, while the obstacles' uncertainty ellipsoids $\sigma_{dH,i}$ and risk levels $\alpha_i$ tighten the constraints under sensing uncertainty. The safe set of action $C$ ensures that the UAV's safe zone is not violated.}
  \label{fig:Avoidance}
\end{figure}

\subsection{Problem Statement}
\label{subsec:PROBLEM}

We model the UAV dynamics by second-order differential equations for simpler control design. We consider an acceleration-controlled quadrotor UAV operating in an environment with $N$ dynamic obstacles. The obstacles in the sensing field of view (FoV) are detected and tracked with typical trackers (e.g., SAFE-IMM tracker~\cite{mandaokar_safeimm}), which deliver state estimates of the obstacles. The input space consists of the position $p_e\in\mathbb{R}^3$, velocity $v_e\in\mathbb{R}^3$, and acceleration $a_e\in\mathbb{R}^3$ of the UAV as well as the relative position, velocity, and acceleration of the obstacle. A nominal controller provides an acceleration command $a_{\mathrm{n}}\in\mathbb{R}^3$, while the UAV is subject to perception-to-actuation latency $\Delta\!\ge\!0$ and bounded actuation, i.e., limits for acceleration $a_{\max}\!>\!0$, jerk $j_{\max}\!>\!0$, and speed $v_{\max}\!>\!0$. The objective of DR-ACBF is to design a real-time safety filter that modifies $a_{\mathrm{n}}$ to a safe command $a_e\in\mathbb{R}^3$ while enforcing probabilistic collision avoidance under a user-specified global risk budget $\alpha_{\mathrm{total}}$ (see Figure~\ref{fig:Avoidance}).

\subsection{Dynamics and Relative Kinematics}
\label{subsec:DYNAMICS}
Similar to the UAV, each dynamic obstacle $i\in\{1,\dots,N\}$ is modeled as 
\begin{equation}
    \dot p_i = v_i,\qquad \dot v_i = a_i,\qquad \|a_i(t)\| \le \hat a_{\max,i},
    \label{eq1}
\end{equation}
where $p_i,v_i,a_i\in\mathbb{R}^3$ denote the position, velocity, and acceleration, and $\hat a_{\max,i}>0$ is a learned upper bound on the obstacle acceleration. 
The relative position and velocity between the UAV and obstacle $i$ are defined as
\begin{equation}
    P_i = p_i - p_e,\qquad V_i = v_i - v_e,
\end{equation}
yielding the relative dynamics
\begin{equation}
    \dot P_i = V_i,\qquad \dot V_i = a_i - a_e.
\label{relposvel}
\end{equation}
From $P_i$ and $V_i$ we define the LoS quantities
\begin{equation}
    n_i = \frac{P_i}{\|P_i\|},\qquad d_i = \|P_i\|,\qquad v_i^{\mathrm{cl}} = \max\{0,\,-n_i^\top V_i\},
\label{relquan}
\end{equation}
where $n_i$ is the LoS unit vector, $d_i$ is the range to the obstacle, and $v_i^{\mathrm{cl}}$ is the closing speed, capturing only approaching components of the relative velocity as illustrated in Figure~\ref{fig:Avoidance}.

\subsection{Tracking Uncertainty and Lookahead Prediction}
\label{subsec:TRACKING}
The state and uncertainty estimates for each obstacle are provided by the tracker
\begin{equation}
    \{\hat p_i,\hat v_i,\Sigma_i,\sigma_{v,i}\}_{i=1}^N,
\end{equation}
where $\Sigma_i\in\mathbb{R}^{3\times 3}$ is the position covariance and $\sigma_{v,i}$ is the standard deviation of the LoS component of the relative velocity uncertainty. The UAV position uncertainty is expressed by $\Sigma_e$. Given a lookahead horizon $H>0$, the LoS distance prediction uncertainty is given as
\begin{equation}
    \sigma_{dH,i} = \sqrt{\,n_i^\top(\Sigma_e+\Sigma_i)n_i + (H\sigma_{v,i})^2\,}.
\label{eq:sigma}
\end{equation}

\subsection{Classical CBF Formulation}
\label{subsec:CBFCLASSICAL}

Ames et al.~\cite{ames2014acc,ames2017tac} introduced the classical CBF-QP architecture, which guarantees that an autonomous agent moves toward its goal while maintaining safety constraints.  
Let $x\in\mathbb{R}^n$ denote the system state and $u\in\mathbb{R}^m$ the control input.  
The control-affine dynamics are
\begin{equation}
\dot{x}=f(x)+g(x)u
\label{eq:affine-classical}
\end{equation}
where $f(x)$ is the drift term and $g(x)$ the control effectiveness matrix. A continuously differentiable function $h:\mathbb{R}^n\!\to\!\mathbb{R}$ defines the safe set $\mathcal{C}$:
\begin{equation}
\mathcal{C} = \{x\in\mathbb{R}^n \mid h(x)\ge0\}.
\label{safeset}
\end{equation}

\noindent The Lie derivatives of $h$ are $L_fh(x)=\nabla h(x)^\top f(x)$ and $L_gh(x)=\nabla h(x)^\top g(x)$.
The function $h$ is a CBF with an extended class-$\mathcal{K}$ function $\alpha(\cdot)$ such that
\begin{equation}
L_fh(x) + L_gh(x)u + \alpha(h(x)) \ge 0.
\label{eq:cbf}
\end{equation}
When~\eqref{eq:cbf} is satisfied (i.e., the CBF constraint is enforced), the safe set $\mathcal{C}$ is forward invariant, meaning $x(0)\in\mathcal{C}$ implies $x(t)\in\mathcal{C}$ for all $t\ge0$. At each control step, one computes $u^\star$ by solving the QP
\begin{equation}
\begin{aligned}
u^\star = \arg\min_{u} \tfrac{1}{2}\|u-u_{\mathrm{nom}}\|^2  \\[2pt]
\text{s.t.}\quad
L_f h(x) + L_g h(x)u + \alpha(h(x)) \ge 0\\
\end{aligned}
\label{eq:clf-cbf-qp-correct}
\end{equation}

\noindent where $u_{\mathrm{nom}}$ is the nominal control. The CBF inequality~\eqref{eq:cbf} strictly enforces the safety required to reach the goal successfully. We omit the control Lyapunov function, as no soft performance constraint is used.

\section{Method}

\begin{figure}
  \centering
  \includegraphics[width=\columnwidth]{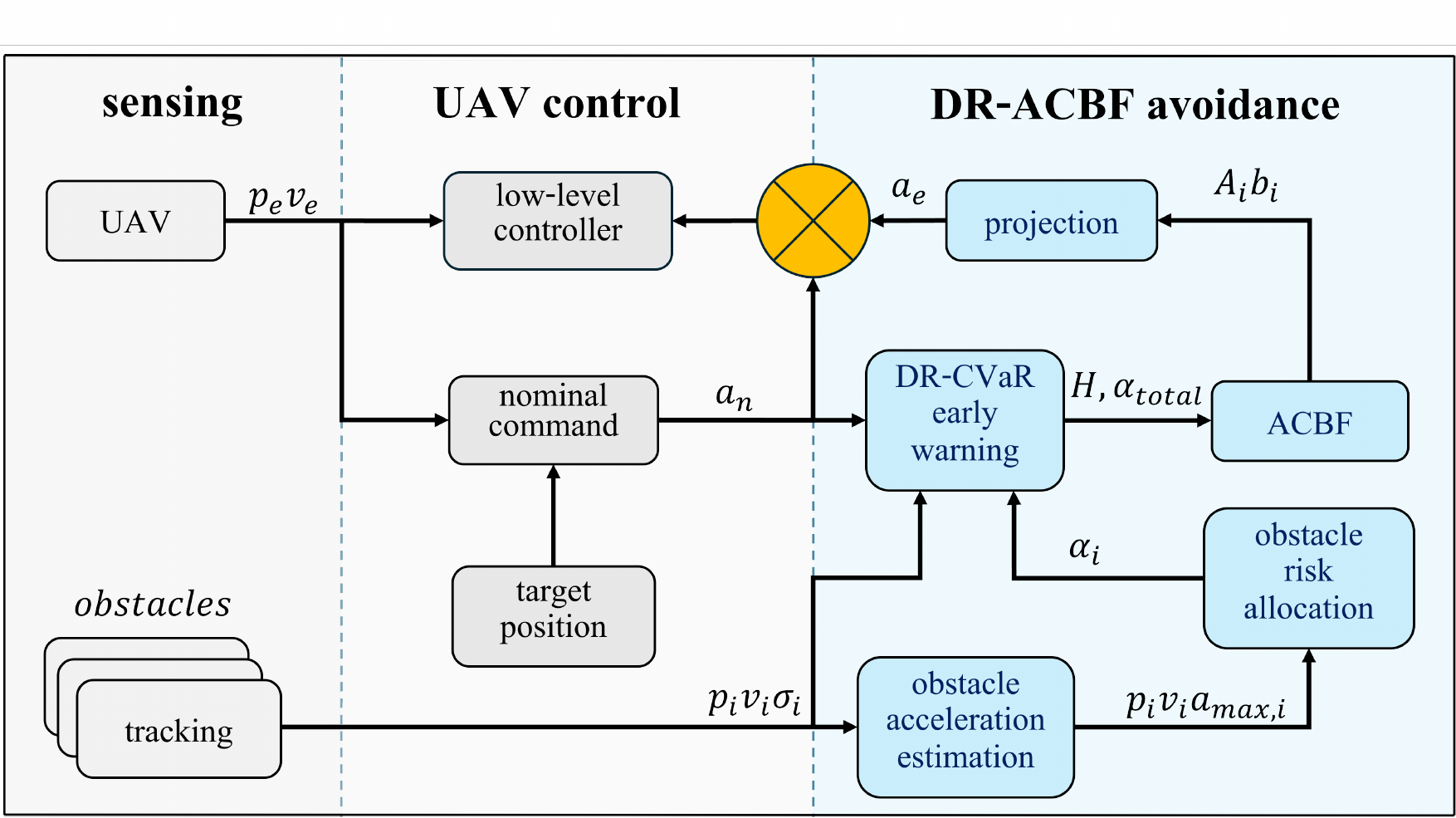}
  \caption{System block diagram. 
The process begins with sensing the UAV and obstacle states. UAV control generates a nominal acceleration command $a_n$, which is corrected by DR-ACBF avoidance. 
The DR-CVaR layer provides an early warning and requires acceleration estimation and risk allocation to compute the safety constraints. Finally, ACBF and projection replace $a_n$ with the safe acceleration command $a_e$ to avoid obstacles.}
  \label{fig:pipeline}
\end{figure}

Figure~\ref{fig:pipeline} shows our system block diagram, where the contribution of this paper (DR-ACBF avoidance) is depicted in blue. This part combines obstacle acceleration estimation and risk association with DR-CVaR to provide an avoidance warning, which serves as input to the ACBF and Gauss-Southwell projection to provide the avoidance acceleration. DR-ACBF does not integrate a trajectory planner; instead, it utilizes the target position as the next lookahead. 

\subsection{Acceleration CBF}
\label{subsec:ACBF}

We model the UAV dynamics similar to Equation~\eqref{eq1} with $\|a_e\|\le a_{\max}$:  
\begin{equation}
\dot{p}_e=v_e,\qquad \dot{v}_e=a_e.
\label{eq:dynamics}
\end{equation}

The combined sum of obstacle and UAV radius is $R_{\mathrm{sum}}=R_{\mathrm{uav}}+R_{\mathrm{obs}}$ as illustrated in Figure~\ref{fig:Avoidance}. The barrier function for distance is then
\begin{equation}
h_i(P_i)=\|P_i\|^2-R_{\mathrm{sum}}^2
\end{equation}
which satisfies $h_i\ge0$ in the safe region.  
Differentiating twice and substituting \eqref{relposvel} gives
\begin{equation}
\dot{h}_i=2P_i^\top V_i,\qquad 
\ddot{h}_i=2V_i^\top V_i+2P_i^\top(a_i-a_e).
\end{equation}

\noindent Therefore, safety is ensured when the second-order barrier condition holds
\begin{equation}
\ddot{h}_i+k_1\dot{h}_i+k_2h_i\ge0,\qquad k_1,k_2>0,
\label{eq:cbf-2nd}
\end{equation}
where $k_1$ damps the approach velocity and $k_2$ penalizes distance violation. In order to project the second-order condition into an acceleration half-space, we linearize it over a short lookahead horizon $H>0$. Where the $a_i$ is not modeled; its worst-case effect is padded into $R_{\mathrm{eff},i}$ and $v_i$ enters via $v_i^{\mathrm{cl}}$, yielding a conservative, closed-form affine constraint enforceable at control rate without a QP in the UAV acceleration
\begin{equation}
A_i a_e \le b_i,\qquad
\label{eq:halfspace-basic}
\end{equation}
\begin{equation}
A_i= n_i^\top,\qquad
b_i=\frac{2}{H^2}\Big(d_i - R_{\mathrm{eff},i} - v_i^{\mathrm{cl}}H\Big)
\label{eq:halfspace}
\end{equation}

\noindent
where $R_{\mathrm{eff},i}$ is the effective clearance that accounts for sensing and actuation latencies. Each obstacle thus defines one linear half-space of safe accelerations. Normalization by $\|P_i\|$ is already included in $n_i$, so~\eqref{eq:halfspace} requires no extra factor.

The effective clearance in distance is enlarged to account for perception latency, actuation limits, and jerk bound
\begin{equation}
R_{\mathrm{eff},i}
=R_{\mathrm{sum}}
+v_{\max}\Delta
+\tfrac{1}{2}a_{\max}\Delta^2
+\tfrac{1}{6}j_{\max}H^3
+\tfrac{1}{2}\hat a_{\max,i}H^2
\label{eq:Reff}
\end{equation}
where $v_{\max}$ is the UAV speed limit, $a_{\max}$ and $j_{\max}$ are acceleration and jerk bounds, $\Delta$ is total perception-to-actuation latency, and the $\hat a_{\max,i}$ is the learned bound on obstacle acceleration. The half-space linearization assumes nearly constant LoS direction and bounded constant acceleration over $H$, and the neglected higher-order effects are conservatively covered by inflating $R_{\mathrm{sum}}$ to $R_{\mathrm{eff}}$ (especially for small $H$). The sliding-mode differentiator (SMD) is used to compute the $\hat a_{\max,i}$ with a user-defined $a_{\max,i}$, aiming to keep it large to address the worst possible acceleration of obstacles. This is explained in detail in the next sections.

To compensate for latency variations, the latency term in $R_{\mathrm{eff}}$ is bounded by $v_{\max}\Delta + \tfrac12 a_{\max}\Delta^2$. While the new command is held over $H$ with bounded jerk $j_{\max}$, the worst additional shrink from the acceleration ramp is upper-bounded by $\tfrac{1}{6}j_{\max}H^3$. The obstacle displacement over $H$ is limited by $\tfrac12\hat a_{\max,i}H^2$. Summing up all these components to $R_{\mathrm{sum}}$ yields \eqref{eq:Reff}. The $\Delta$ includes pre-actuation latency to account for delays in activating the controls. The impact of such latency on safe high-speed navigation with bounded acceleration is analyzed in~\cite{falanga2019ral}. 
The trackers provide estimates of position and velocity~\cite{mandaokar_safeimm}. Therefore, if we consider the position and velocity covariance from the tracker, the LoS variance is bounded by Eq.~\eqref{eq:sigma}. 
Over a short lookahead horizon $H$, the coefficients $k_1=2/H$ and $k_2=2/H^2$ ensure that any acceleration satisfying the affine constraint maintains a non-decreasing barrier value of $h_i(t)$ over $H$. Using the 2$^{nd}$ order Taylor approximation \(h(t+H)\approx h(t)+H\dot{h}(t)+\tfrac{1}{2}H^{2}\ddot{h}(t)\), enforcing \(h(t+H)\ge 0\) yields \(\ddot{h}(t)+\tfrac{2}{H}\dot{h}(t)+\tfrac{2}{H^{2}}h(t)\ge 0\), hence \(k_{1}=\tfrac{2}{H}\) and \(k_{2}=\tfrac{2}{H^{2}}\).

\subsection{Obstacle Risk Allocation}
\label{subsec:RISK}

Risk allocation helps to select the appropriate response when multiple obstacles are present by avoiding uniform conservatism and prioritizing the nearest, fastest-closing obstacle. We allocate a global risk budget $\alpha_{\mathrm{total}}\in(0,1)$ among $N$ obstacles indexed by $i,j \in \{1,2,\ldots,N\}$. We assign obstacle budgets $\alpha_i$ using a proximity-velocity weighting with $v_i^{\mathrm{cl}}$~\eqref{relquan} for some small $\varepsilon > 0$ to regularize the normalization and ensure numerical safety:
\begin{equation}
r_i=\frac{v_i^{\mathrm{cl}}}{\max_j v_j^{\mathrm{cl}} + \varepsilon},\quad
w_i=\frac{1 + r_i}{\max(d_i,10^{-3})},
\label{eq:risk-alloc}
\end{equation}

\begin{equation}
\alpha_i = \frac{w_i}{\sum_j w_j}\,\alpha_{\mathrm{total}}.
\label{eq:risk-alloc_}
\end{equation}
The $10^{-3}$ term in $w_i$ prevents division by zero. The per-obstacle budgets sum to the global risk $\sum_i \alpha_i = \alpha_{\mathrm{total}}$. The $r_i$ is a normalized closing-speed score that prioritizes obstacles approaching faster than others, and $w_i$ is the combined weight for approach urgency and proximity. Therefore, the provided obstacle budgets $\alpha_i$ with a one-sided Cantelli inequality~\cite{Cantelli} yields the tightening factor $\lambda_i$
\begin{equation}
\Pr\!\big( d_{\text{true},i}(t{+}H) \ge d_{\text{pred},i}(t{+}H) - \lambda_i \sigma_{dH,i} \big) \ge 1-\alpha_i,
\label{eq:lambda}
\end{equation}
\begin{equation}
\lambda_i=\sqrt{\frac{1-\alpha_i}{\alpha_i}},
\label{eq:lambda_}
\end{equation}
which allows to tighten the half-space right-hand. The quantity $d_{\text{true},i}(t+H)$ denotes the unknown realized distance at time $t+H$ which is modeled around $d_{\text{pred},i}(t+H)$ for sensing uncertainty as a random variable with standard deviation $\sigma_{dH,i}$ to predict the LoS distance. 
Substituting the uncertainty safety margin into \eqref{eq:halfspace-basic} gives the tightened constraint:
\begin{equation}
A_i a_e \le 
\frac{2}{H^2}\Big(d_i - R_{\mathrm{eff},i} - v_i^{\mathrm{cl}}H - \lambda_i\,\sigma_{dH,i}\Big).
\label{eq:halfspace-tight}
\end{equation}

Temporal safety over $[t,t{+}H]$ is ensured by the clearance $R_{\mathrm{eff},i}$~\eqref{eq:Reff} and the Cantelli tightening~\eqref{eq:lambda}~\eqref{eq:lambda_}. 
The clearance $R_{\mathrm{eff},i}$ expands the safety margin, while the tightening term $\lambda_i\sigma_{dH,i}$ bounds stochastic errors, which yields a closed-form bound for the collision barrier~\eqref{eq:halfspace-tight}. This tightened bound maintains the ACBF inequality~\eqref{eq:cbf-2nd} with probability of at least $1-\alpha_i$. Consequently, the safe set $C$ remains forward invariant under sensing uncertainty and latency.

\subsection{Obstacle Acceleration Estimation}
\label{subsec:OBSACC}

In addition to risk allocation, DOA needs to estimate acceleration bounds for the obstacle to maintain a safe distance in all cases. In order to assess the unknown acceleration of an obstacle $\hat a_{\max,i}$, we assume piecewise constant relative acceleration over the lookahead $H$ and compute the upper bound $\hat a_{\max,i}$ via higher-order SMD~\cite{2nSDM,SDM}. From measured obstacle velocity $v_i$, the SMD internal states $(z_0,z_1)$ yield a robust estimate $\hat a_i=z_1$.
The differentiator dynamics are given as 
\begin{equation}
\begin{split}
\dot{z}_0 &= -l_1\,|z_0 - v_i|^{1/2}\,\mathrm{sgn}(z_0 - v_i) + z_1 \\
\dot{z}_1 &= -l_2\,\mathrm{sgn}(z_0 - v_i) \\
\end{split}
\label{eq:smd-levant}
\end{equation}
where $(z_0,z_1)$ are internal states and $l_1,l_2>0$ are design gains.
The SMD output provides a robust, noise-insensitive estimate of the true derivative $\dot{v}_i = a_i$ with finite-time convergence. The estimated acceleration bound is updated online as
\begin{equation}
\hat{a}_{\max,i} \leftarrow \max\big(\hat{a}_{\max,i},\,\|\hat{a}_i\|\big)
\end{equation}
which ensures a conservative envelope on obstacle acceleration used in the effective clearance~\eqref{eq:Reff} and allows the safety margin to expand automatically when obstacles exhibit more aggressive maneuvers.

\subsection{DR-CVaR Early Warning}
\label{subsec:DRCVaR}
Additionally, the framework uses an early trigger to complement ACBF based on conditional value at risk (CVaR)~\cite{cvar2023,drcvar2023}. We introduce a real-time distributionally robust CVaR (DR-CVaR) layer that operates in conjunction with the proposed ACBF controller. The DR-CVaR trigger samples predictions from the tracker's Gaussian output, assuming constant acceleration and bounded braking actions to check if it violates a dynamic distance to the UAV.
The DR-CVaR anticipates imminent violations and triggers early avoidance. It acts as an early-warning layer that, when activated, tightens ACBF enforcement by adjusting horizon and risk, and only triggers when the risk condition is met (see Fig.~2). Once triggered, the system hands control to the ACBF block, which operates on horizon $H$. For each obstacle $i$ and prediction time $t_k$ sampled on a discrete grid within $ [0,H_{max}]$, the set of propagated sample states is defined by $P^{(n)}(t_k)$ and $V^{(n)}(t_k)$ for the $n$-th sample prediction. For each sample $P^{(n)}, V^{(n)}$, we define the LoS unit vector, i.e., $n^{(n)}$ and the closing speed $v^{(n)}$ computed as in Eq.~\eqref{relquan}. This helps to estimate a dynamic distance for the $n$-th sample
\begin{equation}
D_{\mathrm{dyn}}^{(n)}=v^{(n)}\tau+\frac{(v^{(n)})^2}{2(a_{\mathrm{e}}+a_{\mathrm{i}})}
\label{eq:Ddyn}
\end{equation}
\noindent where $a_{\mathrm{i}}$ and $a_{\mathrm{e}}$  denote the available magnitudes of the obstacle and the UAV deceleration. The $\tau$ denotes the reaction latency.
Then, the safe distance is 
\begin{equation}
D_{\mathrm{safe}}^{(n)}=R_{\mathrm{sum}}+D_{\mathrm{dyn}}^{(n)}.
\end{equation}
The corresponding violation term is
\begin{equation}
Z^{(n)}=D_{\mathrm{safe}}^{(n)}-\|P^{(n)}(t_k)\|.
\end{equation}

To guard against distributional mismatch, we add a Wasserstein upshift $\epsilon L_Z$ with radius $\epsilon$ and Lipschitz constant $L_Z$~\cite{wasserstein}. We adopt the standard empirical CVaR definition using the worst $m$ samples~\cite{cvar2023,drcvar2023}

\begin{equation}
\mathrm{CVaR}_\beta^{\mathrm{DR}}(Z)=\frac{1}{m}\sum_{n\in\text{worst }m} Z^{(n)}+\epsilon L_Z
\label{eq:DRcvar}
\end{equation}
\noindent
where $m = \lceil \beta N_s \rceil$ with  $\beta\in(0,1)$
specifies how much of the worst-case distribution we consider. $N_s$ denotes the number of Monte-Carlo samples per obstacle used for short-horizon prediction. We set $\epsilon$ as a robust radius tied to the sample spread. If $\max_k\mathrm{CVaR}_\beta^{\mathrm{DR}}(Z)\ge0$, the early warning layer triggers avoidance before any barrier violation is imminent. The trigger activates ACBF with a projection to replace the nominal acceleration $a_\mathrm{n}$ by an evasive avoid acceleration $a_\mathrm{e}$. It tightens the ACBF constraints and $R_{eff}$ by reducing $\alpha_\mathrm{total}$ and increasing $H$ for one cycle. The $\beta$ controls the tail fraction for prediction-based early warning, and the $\alpha_\mathrm{total}$ provides the probabilistic tightening of the ACBF constraints.

\subsection{Gauss-Southwell Projection}
\label{subsec:GUASS}

To improve the real-time capabilities, we avoid QP and apply a deterministic Gauss-Southwell (most violated constraint) projection to a nominal acceleration $a_0 = a_{\mathrm{n}}$~\cite{fastproj,gauss}
\begin{equation}
\begin{aligned}
i^\star &= \arg\max_i(A_i a_l-b_i)
\end{aligned}
\end{equation}
where $i^\star$ is the index of the most violated constraint and $a_l$ is the acceleration iterate at the current step of the projection loop. At each iteration $l$, the acceleration is updated as

\begin{equation}
\begin{aligned}
a_{l+1} &= a_l-\omega\frac{(A_{i^\star} a_l-b_{i^\star})_+}{\|A_{i^\star}\|^2}A_{i^\star}^\top \\
a_{l+1}&\leftarrow \Pi_{\|a\|_\infty\le a_{\max}}(a_{l+1})
\end{aligned}
\label{eq:proj}
\end{equation}
\noindent
where $\omega\in(0,1]$ is a relaxation coefficient in the projection algorithm, which is tuneable empirically for fastest convergence. The resulting $a_e=a_l$ is the safe acceleration command. The iteration \eqref{eq:proj} is a relaxed Gauss-Southwell projection onto the most violated halfspace, followed by projection onto the acceleration space~\eqref{eq:halfspace-basic}. In addition, the loop can stop early when no constraint is violated, i.e., when $\max_i(A_i a_l-b_i)\le 0$. Gauss-Southwell projection is known to be numerically stable~\cite{fastproj,gauss} and fast by limiting the iterations. The projector iterates to satisfy~\eqref{eq:halfspace-tight} and reach the feasible set $C$~\eqref{safeset}. We run a fixed number of iterations and check the maximum violation; if it exceeds a threshold, a fallback is triggered by increasing $H$ and tightening the risk budget. If violations persist, the projector biases the command toward braking and landing.

\section{Simulation Study}
\label{sec:sim}
In this section, we use the classical velocity-based CBF-QP~\cite{ames2014acc,AdaptiveCBF2022}, robust policy CBF (RP-CBF)~\cite{SafetyFly2025}, and model predictive control (MPC)~\cite{lindqvist2020ral,ampc,tinyMPC} as baselines and integrate the identical DR-CVaR early trigger on all approaches for a fair comparison of the ACBF improvement. For the MPC baseline~\cite{mpc,ampc}, we use MATLAB's built-in MPC (QP solver) with prediction horizon $p=20$ and control horizon $m=3$ for real-time constraints. For RP-CBF~\cite{SafetyFly2025}, horizon $T_{\text{pcbf}}=3.0$\,s and $N=32$ disturbance samples are incorporated via tracker's obstacle-motion estimation. The CBF-QP uses the RP-CBF configuration and horizon of $0.4$\,s.

We use the following metrics for our comparison: the success rate $s$ is defined as the ratio of simulation runs that successfully avoid collisions over all runs. The computation time $t_{cp}$ is the processing time of our avoidance algorithm (from acceleration estimation to projection in Figure~\ref{fig:pipeline}). The command time $t_{cm}$ represents the control latency, defined as the time from the trigger to issuing the first avoid command. The reaction time $t_{r}$ represents the duration from the trigger to the first deviation from the nominal command and combines the $t_{cm}$ with the dynamics of the UAV. The minimum separation $d_s$ represents the minimum distance between the UAV and the obstacle during the avoidance maneuver. Finally, $v_{cl}$ represents the speed of the obstacle towards the UAV.

\subsection{Simulation Setup}
\label{subsec:simsetup}

We simulate the proposed method in MATLAB on a CPU-only laptop using the UAV Scenario Toolbox~\cite{UAVtoolbox} at $100$\,Hz. The dynamic obstacles move across the LoS with random initial velocities, with a lower bound of $v_{\min}=5$\,m/s. The UAV and obstacle radius are $R_{\mathrm{uav}}=0.15$\,m and $R_{\mathrm{obs}}=0.15$\,m resulting in $R_{\mathrm{sum}}=0.3$\,m. The actuation limits are bounded with $\|a_e\|_\infty\le6$\,m/s$^2$. A 3D RADAR sensor data generator~\cite{radarDataGenerator} with 80\,azimuth$\times$20$\,$elevation, and range of 150$\,$m is used to define the FoV (as depicted in Figure~\ref{fig:Avoidance}) and generates realistic mmWave RADAR data for 
the GNN-based SAFE-IMM obstacle tracker~\cite{mandaokar_safeimm}. 

DR-ACBF uses a one-step global collision risk budget of $\alpha_{\mathrm{total}}=0.10$, distributed among obstacles according to proximity $d_i$ and closing speed $v^{\mathrm{cl}}_i$. A single horizon per cycle $H$ is selected via coarse search $H\in[H_{\min},H_{\mathrm{max}}]$ with user given time horizon $H_{\mathrm{max}}=0.4$\,s and $H_{\min}=0.15$\,s. 
The coarse search selects a feasible $H$ that maximizes clearance and satisfies all constraints. Obstacle acceleration envelopes $\hat a_{\max,i}$ are learned online using an SMD differentiator with $(\gamma,L_0,L_1)=(1.5,4,3)$. Safe accelerations are obtained via a fixed-time Gauss-Southwell projector with 15 iterations and relaxation $\omega=1.0$. The complete DOA consists of the avoidance action, followed by a hover at the bypass position, and mission continuation towards the next target position. For avoidance direction, the best of 12 candidate directions (axis/diagonal) for each step is selected to minimize lateral acceleration over $H$, providing a bypass position with clearance offset. The UAV should hover with time to stop within $t_{\mathrm{s}}=0.5$\,s and braking intensity of $k_{\mathrm{s}}=1.0$. The lateral clearance $d_{cl}$ is $1.0$\,m, and minimum hold time after avoid $t_{hd}$ is 0.2\,s. The DR-CVaR early-warning trigger runs before projection with $\beta=0.05$ and a reaction latency of $\tau=0.02$\,s. We set the Wasserstein radius $\epsilon$ to 0.05 for numerical stability.

\subsection{Simulation Results}

This section presents the results for the first scenario, where the trajectory of one obstacle is curving to intersect the UAV's path, and the other obstacle moves directly towards the UAV. Both obstacles start at 50\,m from the UAV with small detection noise and increasing speed. Table~\ref{tab:scene1-sim} compares the proposed DR-ACBF approach against conventional CBF, MPC, and recent RP-CBF techniques with identical simulation scenarios of two dynamic obstacles. Our DR-ACBF outperforms the other techniques in all time measurements. Particularly, the processing time is reduced by 31\% to 54\% with lower variability, which is important for embedded platforms. Despite the reduced computational effort, DR-ACBF maintains comparable performance with a mean success rate $s$ of 99.64\% and the minimum separation $d_{s}$ of 4.73\,m, satisfying safety constraints throughout the mission. Standard deviations reflect variability for methods (with extreme outliers removed); overall low for DR-ACBF and high $t_{r}$ RP-CBF’s due to obstacle 1.

\begin{table}[H]
   \centering
   \caption{Comparison of DR-ACBF with three baseline DOA methods using 3000 simulations for each scenario.}
   \label{tab:scene1-sim}
   \setlength{\tabcolsep}{2pt}
   \renewcommand{\arraystretch}{1.1}
   \begin{tabular}{|c|c|c|c|c|c|c|c|}
     \hline
     \multirow{2}{*}{\textbf{Method}} &
     \multirow{2}{*}{\textbf{ Obstacle }} &
     $s$ & $v_{cl}$ & $t_{r}$ & $t_{cm}$ & $t_{cp}$ & $d_{s}$ \\
     & & (\%) & (m/s) & (ms) & (ms) & (ms) & (m) \\ \hline
     \multirow{3}{*}{MPC} & 1  & 99.52 & 10.43 & 10.14 & 3.23 & - & 2.82\\ 
                       & 2  & 99.89 & 10.97 & 10.00 & 3.21 & - & 6.64 \\ 
                       &  \textbf{Mean} & \textbf{99.71} & \textbf{10.70} & \textbf{10.07} & \textbf{3.22} & \textbf{0.956}& \textbf{4.73} \\
                       &  \textbf{Std} & \textbf{-} & \textbf{-} & \textbf{4.52} & \textbf{0.93} & \textbf{0.666} & \textbf{-} \\ \hline
     \multirow{3}{*}{CBF-QP} & 1  & 99.51 & 10.43 & 10.03 & 2.21 & - & 2.82\\ 
                       & 2  & 99.84 & 11.00 & 10.00 & 2.19 & - & 6.65 \\ 
                       &  \textbf{Mean} & \textbf{99.67} & \textbf{10.71} & \textbf{10.02} & \textbf{2.20} & \textbf{0.714} & \textbf{4.74} \\
                       &  \textbf{Std} & \textbf{-} & \textbf{-} & \textbf{1.31} & \textbf{0.46} & \textbf{0.476} & \textbf{-} \\ \hline 
     \multirow{3}{*}{RP-CBF} & 1  & 99.52 & 10.43 & 106.26 & 1.87 & - & 2.82\\ 
                       & 2  & 99.89 & 10.97 & 13.39 & 2.26 & - & 6.64 \\ 
                       &  \textbf{Mean} & \textbf{99.71} & \textbf{10.70} & \textbf{59.83} & \textbf{2.08} & \textbf{0.642} & \textbf{4.73} \\ 
                       &  \textbf{Std} & \textbf{-} & \textbf{-} & \textbf{93.48} & \textbf{0.48} & \textbf{0.416} & \textbf{-} \\ \hline 
     \multirow{3}{*}{DR-ACBF} & 1  & 99.48 & 10.43 & 10.06 & 1.39 & - & 2.82\\ 
                       & 2  & 99.80 & 10.98 & 10.00 & 1.41 & - & 6.64 \\ 
                       &  \textbf{Mean} & \textbf{99.64} & \textbf{10.70} & \textbf{10.03} & \textbf{1.40} & \textbf{0.444} & \textbf{4.73} \\
                       &  \textbf{Std} & \textbf{-} & \textbf{-} & \textbf{2.76} & \textbf{0.38} & \textbf{0.270} & \textbf{-} \\ \hline 
   \end{tabular}%
 \end{table}

\subsection{Sensitivity Analysis and Ablation }

We assessed the robustness of the proposed method and the tracker's impact via a sensitivity analysis and ablation study with 1000 runs in a second scenario, varying three key parameters. The second scenario has one obstacle crossing diagonally and another moving straight toward the UAV, with initial distances of 150$\,$m and 200$\, $m. This scenario tests linear-motion avoidance under higher long-range detection noise, affecting tracking and reducing success. Table~\ref{tab:sen} summarizes the sensitivity of DR-ACBF to the key parameters maximum time horizon $H_{\mathrm{max}}$, clearance distance $d_{cl}$, and hold time $t_{hd}$. Each parameter was varied independently while keeping the others fixed. Results indicate that DR-ACBF maintains a consistent $s$ above 77\% in this scene and shows no significant change in $t_{r}$ and $d_{s}$ across all key parameter configurations. Minor variations in $H_{\mathrm{max}}$ show a reduction in $t_{cm}$ and $t_{cp}$. 
\begin{table}[H]
  \centering
  \caption{Sensitivity analysis of key DR-ACBF Parameters averaged for 1000 simulation runs.}
  \label{tab:sen}
  \setlength{\tabcolsep}{2pt}
  \renewcommand{\arraystretch}{1.1}
  \begin{tabular}{|c|c|c|c|c|c|c|c|c|c|}
    \hline
$H_{\mathrm{max}}$ & $d_{cl}$ & $t_{hd}$ & $s$ & $v_{cl}$ & $t_r$ & $t_{cm}$ & $t_{cp}$ & $d_s$ \\
(s) & (m) & (s) & (\%) & (m/s) & (ms) & (ms) & (ms) & (m) \\
    \hline
    
    0.4 & 3 & 0.2 & 82.5 & 8.18 & 12.74 & 2.24 & 0.475 & 1.881 \\ 
    0.2 & 3 & 0.2 & 79.8 & 8.16 & \textbf{11.88} & \textbf{1.58} & 0.312 & 1.822 \\
    0.8 & 3 & 0.2 & 82.5 & 8.28 & 12.35 & 1.85 & 0.382 & 2.105 \\ 
    0.4 & 1 & 0.2 & 82.3 & 8.02 & 12.87 & 2.07 & \textbf{0.234} & 1.864 \\
    0.4 & 4 & 0.2 & 77.6 & 7.16 & 12.16 & 1.66 & 0.423 & \textbf{2.054} \\ 
    0.4 & 3 & 0.1 & \textbf{82.6} & 7.10 & 11.92 & 1.72 & 0.362 & 1.924 \\
    0.4 & 3 & 0.4 & 82.6 & 8.18 & 12.17 & 1.97 & 0.403 & 1.945 \\ \hline                     
  \end{tabular}%
\end{table}

Finally, we conduct an ablation study to evaluate the effectiveness of the proposed early trigger DR-CVaR in ACBF. We compared it with a baseline, i.e., a simple trigger that uses a single-time Gaussian chance constraint by evaluating relative distance at that single time (DR-CVaR OFF in Table~\ref{tab:abl}). The simple trigger is required as the ACBF behaves like a reactive safety filter; without the trigger, it may activate late under latency, even with deterministic settings.  All other components (geometry, dynamics and parameters as in first row of Table~\ref{tab:sen}) remain identical across 1000 runs. Table~\ref{tab:abl} shows that DR-CVaR ON leads to a substantial improvement in performance, increasing the success rate $s$ from 34.9\% to 82.5\% while maintaining comparable response times. This highlights the need for an early trigger; the adaptive DR-CVaR ensures timely ACBF activation, whereas a simple trigger leads to late or no ACBF activation, significantly reducing the success rate.
\begin{table}[H]
  \centering
  \caption{Ablation study of DR-CVaR trigger in ACBF for 1000 runs.}
  \label{tab:abl}
  \setlength{\tabcolsep}{2pt}
  \renewcommand{\arraystretch}{1.1}
  \begin{tabular}{|c|c|c|c|c|c|c|c|}
    \hline
    \multirow{2}{*}{\textbf{DR-CVaR}} & \multirow{2}{*}{\textbf{Obstacle}} & $s$ & $v_{cl}$ & $t_r$ & $t_{cm}$ & $d_s$ \\
     & & (\%) & (m/s) & (ms) & (ms) & (m) \\
    \hline 
    \multirow{3}{*}{ON} & 1  & 80.0 & 9.48 & 12.11 & 2.11 & 2.969 \\ 
                      & 2  & 84.7 & 7.10 & 12.95 & 2.35 & 0.985 \\ 
                      &  \textbf{Mean} & \textbf{82.5} & \textbf{8.18} & \textbf{12.74} & \textbf{2.24} & \textbf{1.881} \\ \hline 
    \multirow{3}{*}{OFF} & 1   & 32.0 & 9.70 & 11.74 & 1.74 & 0.707 \\ 
                      & 2  & 37.9 & 8.01 & 12.06 & 2.06 & 0.858 \\ 
                      & \textbf{Mean} & \textbf{34.9} & \textbf{8.70} & \textbf{11.93} & \textbf{1.93} & \textbf{0.796} \\ \hline                     
  \end{tabular}
\end{table}

 The separation distance $d_{s}$ improves from 0.796\,m to 1.881\,m, indicating enhanced safety margins and more reliable avoidance behavior. These results clearly demonstrate that the DR-CVaR trigger is essential for improving safety. Note that the success rate of ACBF can also be improved with DR-CVaR OFF by increasing both $H_{\min}$ and $H_{\max}$, but at the cost of slower reaction time. Thus, the DR-CVaR trigger enables fast reaction. We compared projection-based and QP-based DR-ACBF with the same setup. Success rates were similar, but QP-based had larger mean $t_{cp}$ and $t_{cm}$ (0.766\,ms, 5.36\,ms) than projection-based DR-ACBF (0.475\,ms, 2.24\,ms).

\section{Hardware Experiments}

\subsection{Experimental Setup}
\label{subsec:hwsetup}

The hardware experiments demonstrate the real-time performance of our DR-ACBF DOA and are based on a Bitcraze Crazyflie 2.1 UAV equipped with a flow deck sensor\footnote {\url{https://www.bitcraze.io/}} and an onboard fly unit.  
Contrary to the simulation study with long flight distances, hardware experiments were limited by our OptiTrack system's perception area of about 2\,m$\times$~3\,m (\url{https://youtu.be/CpgcSYkyNRk}). OptiTrack samples the UAV's and obstacles' position at 120\,Hz, which are tracked by the GNN-based SAFE-IMM. We considered key Crazyflie limitations for our setup: size, actuation bounds (acceleration, braking, maneuvers), system latency, and clearance requirements. We have adopted a similar metric for evaluation as used for our simulation study, and additionally measure the avoid time $t_{a}$ and the detection time $t_{d}$. The avoid time is the duration from the start of the avoid action to the end of hovering. The detection time is the duration from the first detection of the obstacle until the avoidance command is issued. A Linux-based CPU-only base station is used to acquire data from OptiTrack, integrate it into MATLAB via a Python bridge, and communicate with the UAV via the Crazyflie MATLAB-Python API at 100\,Hz using full-state setpoints (position, velocity, acceleration, and attitude). The hardware setup uses the identical DR-ACBF configuration as in the simulation study (see Section~\ref{subsec:simsetup}), except for $\|a_e\|_\infty \le 0.8\,\mathrm{m/s^2}$, $R_{\mathrm{uav}}=0.10$\,m and $R_{\mathrm{obs}}=0.10$\,m. Furthermore, $d_{cl}$ is set to $0.05$\,m and $t_{hd}$ to 0.2\,s with the braking configuration: time to stop $t_{\mathrm{s}}$ $\leq$ 0.7\,s and braking intensity $k_{\mathrm{s}}=0.2$. 

\subsection{Experimental Results}

\begin{table}[H]
  \centering
  \caption{Crazyflie avoiding one dynamic obstacle with DR-ACBF in UAV hovering and moving forward condition.}
  \label{tab:flight_metrics}
  \setlength{\tabcolsep}{3pt}
  \renewcommand{\arraystretch}{1.1}
  \begin{tabular}{|c|c|c|c|c|c|c|c|}
    \hline
    \multirow{2}{*}{\textbf{Condition}} & \multirow{2}{*}{\textbf{Flight}} & $v_{cl}$ & $t_r$ & $t_{cm}$ & $t_{a}$ & $t_{d}$ & $d_s$ \\
    & & (m/s) & (ms) & (ms) & (ms) & (ms) & (m) \\
    \hline
    \multirow{5}{*}{Hovering} & 1  & 2.26 & 50.0 & 39.7 & 232 & 3866 & 0.205 \\ 
                      & 2  & 2.56 & 66.5 & 61.5 & 283 & 3617 & 0.863 \\ 
                      & 3  & 3.44 & 66.8 & 59.0 & 233 & 2999 & 0.393 \\ 
                      & 4  & 3.66 & 66.9 & 60.7 & 233 & 3049 & 0.239 \\ 
                      & 5  & 3.60 & 82.8 & 80.7 & 216 & 2516 & 0.695 \\ 
                      &  \textbf{Mean} & \textbf{3.10} & \textbf{66.6} & \textbf{60.3} & \textbf{239} & \textbf{3209} & \textbf{0.479} \\ \hline 
    \multirow{5}{*}{Moving} & 1  & 3.50 & 84.2 & 74.5 & 218 & 2599 & 0.312 \\ 
                      & 2  & 3.74 & 84.0 & 71.6 & 241 & 2725 & 0.258 \\ 
                      & 3  & 3.49 & 67.2 & 52.3 & 208 & 2616 & 0.446 \\ 
                      & 4  & 3.36 & 82.9 & 70.3 & 225 & 2923 & 0.414 \\ 
                      & 5  & 2.90 & 75.6 & 63.1 & 226 & 2678 & 0.324 \\ 
                      & \textbf{Mean} & \textbf{3.40} & \textbf{78.8} & \textbf{66.4} & \textbf{223} & \textbf{2708} & \textbf{0.351} \\ \hline 
  \end{tabular}%
\end{table}

We conduct three hardware experiments. In the first two experiments, the UAV hovers at a fixed position and moves toward a waypoint, respectively. In both experiments, it avoids a single obstacle thrown by a human. In the third experiment, the hovering UAV avoids two obstacles: a directly approaching UAV and a human-thrown object.
Table~\ref{tab:flight_metrics} presents the results of 5 flights of the first and second experiment using our DR-ACBF framework.
In all flights, the UAV successfully avoided the thrown obstacle, which had a closing velocity $v_{cl}$ between 2.26\,m/s and 3.74\,m/s. 
The reaction time $t_{r}$ was shorter in the hovering ($66.6$\,ms) than in the moving ($78.8$\,ms) experiment. The mean command time  $t_{cm}$ showed a similar trend  ($60.3$\,ms vs.~$66.4$\,ms). Whereas the mean avoidance duration $t_{a}$ and detection times $t_{d}$ were smaller for the moving experiment ($223$\,ms and $2.7$\,s vs.~$239$\,ms and $3.2$\,s). This suggests a reliable trigger, as the intervention occurs only when the obstacle becomes unsafe.
The separation distance $d_s$ between the UAV and the obstacle was always greater than the specified $R_{sum}=0.2$\,m, with a mean of $0.479$\,m for hovering and $0.351$\,m moving experiment, ensuring collision-free operation within the tested environment. The simulation $t_{cm}$ is smaller than the hardware due to conservative safety settings. Hardware presents additional delays due to communication and actuation limitations imposed by low battery levels.

\begin{table}[H]
  \centering
  \caption{Crazyflie avoiding two dynamic obstacles with DR-ACBF.}
  \label{tab:flight_metrics1}
  \setlength{\tabcolsep}{3pt}
  \renewcommand{\arraystretch}{1.1}
  \begin{tabular}{|c|c|c|c|c|c|c|c|}
    \hline
    \multirow{2}{*}{\textbf{Obstacle}} & \multirow{2}{*}{\textbf{Flight}} & $v_{cl}$ & $t_r$ & $t_{cm}$ & $t_{a}$ & $t_{d}$ & $d_s$ \\
    & & (m/s) & (ms) & (ms) & (ms) & (ms) & (m) \\
    \hline
    \multirow{5}{*}{1} & 1  & 1.99 & 82.8 & 56.0 & 233 & 1884 & 0.201 \\ 
                      & 2  & 2.37 & 58.2 & 39.4 & 208 & 421 & 0.486 \\ 
                      & 3  & 2.86 & 59.1 & 34.7 & 184 & 342 & 0.281 \\ 
                      & 4  & 3.10 & 49.5 & 36.9 & 158 & 374 & 0.390 \\ 
                      & 5  & 1.43 & 58.1 & 40.5 & 116 & 375 & 0.267 \\ 
                      &  \textbf{Mean} & \textbf{2.35} & \textbf{61.5} & \textbf{41.5} & \textbf{180} & \textbf{671} & \textbf{0.325} \\ \hline 
    \multirow{5}{*}{2} & 1  & 1.21 & 141.6 & 113.2 & 216 & 5050 & 0.657 \\ 
                      & 2  & 1.25 & 80.6 & 62.1 & 214 & 5167 & 0.780 \\ 
                      & 3  & 1.44 & 83.3 & 68.8 & 241 & 3766 & 0.550 \\ 
                      & 4  & 1.74 & 92.3 & 68.5 & 200 & 3732 & 0.738 \\ 
                      & 5  & 1.80 & 99.3 & 80.7 & 216 & 3517 & 0.702 \\ 
                      & \textbf{Mean} & \textbf{1.49} & \textbf{99.4} & \textbf{78.7} & \textbf{217} & \textbf{4246} & \textbf{0.685} \\ \hline 
  \end{tabular}%
\end{table}

Table~\ref{tab:flight_metrics1} presents the results of the third experiment, where the UAV successfully avoided the two obstacles in all five flights.
For the approaching UAV (obstacle 1), the average closing velocity $v_{cl}$ was 2.35\,m/s. We measured similar reaction and command times as in the first experiment (mean values for $t_{r}$ of 61.5\,ms and for $t_{cm}$ of 41.5\,ms). The avoidance duration was on average 0.18\,s, while the mean detection time $t_{d}$ was 0.67\,s. The mean separation distance $d_{s}$ was 0.325\,m, confirming safe maneuvering performance. For obstacle 2, which involved a more unpredictable human-thrown object, the average $v_{cl}$ decreased to 1.49\,m/s. The mean reaction and command times $t_{r}$ and $t_{cm}$ increased to 99.42\,ms and 78.66\,ms, respectively, while the average avoidance duration $t_a$ slightly increased to 0.217\,s. The mean detection time $t_{d}$ significantly increased to 4.25\,s, indicating that the object was detected early and entered the avoidance trigger after some time. Despite these changes, the UAV maintained a safe separation $d_{s}$ of 0.685\,m from obstacle~2. Finally, an important finding of the hardware experiments was that the separation distance $d_{s}$ was above the $R_{\mathrm{sum}}$ bound of  0.2\,m. DR-ACBF was able to satisfy the specified safety constraint for all flights.

\section{Conclusion}

In this paper, we presented a distributionally robust acceleration-level safety filter that enforces collision avoidance through linear, half-space constraints on the commanded acceleration. Our DR-ACBF attains real-time avoidance without solving QPs. Simulation studies against three baseline methods demonstrate that DR-ACBF achieves comparable success rates and safety margins, while reducing computation and command times, thereby improving responsiveness for fast DOA. The sensitivity analysis confirms robustness to key design parameters, with high success rates. An ablation study further demonstrates that the DR-CVaR trigger is crucial for risk-aware operation, and projection is much lighter than QP-based ACBF. Crazyflie experiments demonstrate that DR-ACBF achieves reliable DOA on resource-limited platforms, bridging the gap between formal safety and practical real-time performance. Future work will extend this framework to integrate RADAR or vision sensing modality with a medium-sized UAV to assess the applicability of the proposed framework.












\balance
\bibliographystyle{IEEEtran}
\bibliography{main}

\end{document}